# Transport Properties of "Extended-s" State Superconductors

L.S. Borkowski*, P.J. Hirschfeld* and W.O. Putikka**

*Dept. of Physics, Univ. of Florida, Gainesville, FL 32611, USA
**National High Magnetic Field Laboratory, Florida State University, Tallahassee, FL 32306

Superconducting states with "extended s-wave" symmetry have been suggested in connection with recent ARPES experiments on BSCCO. In the presence of impurities, thermodynamic properties of such states reflect a residual density of states $N(0)$ for a range of concentrations. While properties reflecting $N(\omega)$ alone will be similar to those of d-wave states, transport measurements may be shown to qualitatively distinguish between the two. In contrast to the d-wave case with unitarity limit scattering, limiting low-temperature residual conductivities in the s-wave state are large and scale inversely with impurity concentration.

*Introduction.* Recent high resolution angle-resolved photoemission (ARPES) experiments on Bi-2212 have been interpreted in terms of a highly anisotropic order parameter with a large gap in the $(\pi, 0)$ direction and two lines of nodes near a small gap in the $(\pi, \pi)$ direction. [1] While alternative explanations consistent with an order parameter $\Delta_k$ which reduces the symmetry of the Fermi surface (e.g. d-wave, see Figure 1a) have been put forward, [2] it is interesting to consider the consequences of the simpler suggestion that $\Delta_k$ has the full symmetry of the crystal, but changes sign (see Figure 1b). Fehrenbacher and Norman [3] have considered the effects of potential scattering on such a state, following earlier work on anisotropic-s states with nodes, [4,5] and shown that small concentrations of impurities can lead to "gapless" behavior in the density of states (i.e., with residual density of states $N(0) > 0$), followed by the opening of an actual induced gap in the quasiparticle spectrum as the concentration is increased further.

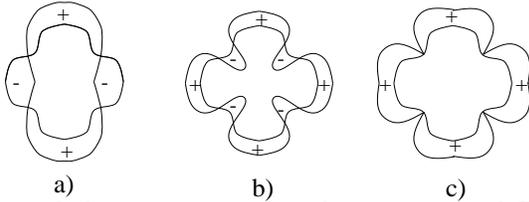

FIG. 1. Order parameters plotted over tetragonal Fermi surface: a) $d_{x^2-y^2}$ state. b) extended s-wave state; c) s-wave state with gap minima at Fermi surface.

The prediction of a range of impurity concentrations over which "gapless" behavior is predicted is important because microwave and NMR experiments, particularly on Zn and Ni-doped YBCO crystals [6,7] have shown evidence of low-temperature thermodynamic properties reflecting the existence of a residual density of states. They have been interpreted most often in terms of a d-wave pairing scenario, [8–10] where the well-known effect of dirt is to induce a residual density of states for infinitesimal concentrations. [11] The intriguing possibility raised by the ARPES experiment [1] is that many of the observations of "gapless" behavior can be equally well explained by an anisotropic state with extended-s symmetry. We propose here to examine this proposition critically.

Before beginning, it should be noted that the scenario suggested by the ARPES work of Ref. 1 faces several difficulties even if it can successfully account for the microwave and NMR data. The most important of these is the set of SQUID experiments [12] indicating that the order parameter $\Delta_k$ changes sign under a $\pi/2$ rotation. While the extended-s order parameter discussed here indeed changes sign over the Fermi surface, the symmetry of the expected tunneling currents is not, within the simplest theory, consistent with the observations reported. [13,14] Furthermore, the existence of a small gap in the $(\pi, \pi)$ direction in BSCCO has been questioned by at least one other photoemission group claiming similar angular and energy resolution. [15] We do not address any of these discrepancies in this work.

In the absence of definitive answers to the above questions, we assume the plausibility of the Argonne-U. Illinois argument and investigate the consequences of assuming the order parameter symmetry identified in Ref. [1] within a generalized BCS model. While these authors pointed out that their data were consistent with an order parameter over a realistic BSCCO Fermi surface with symmetry $\Delta_k \sim \cos k_x \cos k_y$, we work here with an even simpler model with cylindrical Fermi surface, in which case a rather good fit to the data can be obtained by assuming $\Delta_k = \Delta_0(|\cos 2\phi| - \eta_0)$, with $\Delta_0 = 35$meV and $\eta_0 = 0.25$, as shown in Figure 2. With this set of parameters, a BCS weak-coupling approach yields the gap magnitude ratio $\Delta_0/T_c = 2.92$. Note that $\eta_0$ controls the angular range over which the s-wave order parameter is negative. The case $\eta_0 = 0$ corresponds to a somewhat pathological state of type c), while $\eta_0 = 2/\pi$ corresponds to equal positive and negative weights, $\langle \Delta_k \rangle = 0$, where $\langle ... \rangle$ is a Fermi surface average. Other choices of basis functions may result in a somewhat better overall fit and avoid the cusp at $\phi = \pi/4$, but we do not expect these details to affect our qualitative conclusions.

*Consequences of order parameter symmetry.* It is sometimes stated that measurements of penetration depth, angle-resolved photoemission, thermal conductivity, and nuclear magnetic relaxation experiments provide evidence for gap nodes, but do not determine if the order parameter changes sign. This is because experiments of this kind are normally assumed to measure properties sensitive to the order parameter $\Delta_k$ only through the Bogoliubov quasiparticle spectrum, $E_k = \sqrt{\xi_k^2 + |\Delta_k|^2}$. Since $E_k$ does not depend on the sign of the order parameter, all three states shown in Figure 1 should have the same (linear) low-energy density of single-particle states $N(\omega)$, and all properties deriving directly therefrom.

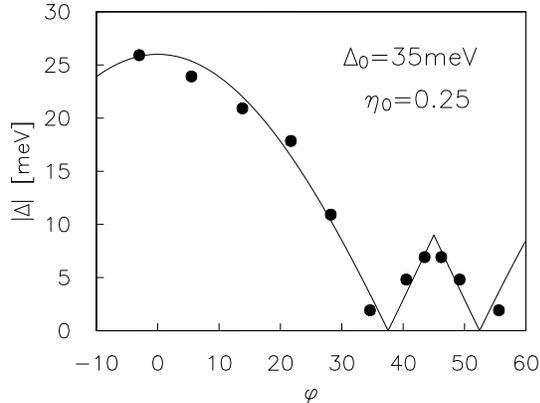

FIG. 2. ARPES determination of BSCCO energy gap as function of angle $\phi$ around Fermi surface. Data from Ref. 1. Solid line: $\Delta_k = \Delta_0(|\cos 2\phi| - \eta_0)$, with $\eta_0 = 0.25$.

Systematic impurity-doping studies of transport properties can in principle help to identify the order parameter symmetry, however. Unconventional states like the $d_{x^2-y^2}$ state (Figure 1a) possess nodes on the Fermi surface for symmetry reasons, whereas their existence is "accidental" in the s-wave case (Figure 1b-c). In the d-wave case, supression of gap variation in **k**-space can only result in an overall supression of the order parameter magnitude, whereas in s-wave cases b) and c), it results eventually in the elimination of the nodes. In many-body language, the crucial point is that the off-diagonal impurity self-energy $\Sigma_1$ is nonzero in s-wave states 1a and 1b, but vanishes in the d-wave state 1c. In the isotropic s-wave state, a large $\Sigma_1$ cancels the diagonal self-energy $\Sigma_0$, [16] leading to Anderson's theorem, [17] whereas in extended-s states this cancellation is only partial. A large $\Sigma_1$ furthermore prevents the formation of a scattering resonance at the Fermi surface, leading to clean limit low-frequency transport coefficients which in extended-s states for $T \ll T_c$ vary relatively weakly with temperature. By contrast, in the d-wave (or other unconventional state where $\Sigma_1 = 0$), resonant scattering leads to transport properties which vary strongly in temperature for $T \ll T_c$. [18–20] In this sense the characteristics attributed to extended-s states here are also valid qualitatively for any mechanism which generates an off-diagonal self-energy in a d-wave state as well, e.g. tetragonal symmetry breaking or locally induced s-wave components due to impurity potentials.

*Single-particle properties.* The disorder-averaged matrix propagator describing any of the states a)-c) above is written

$$\underline{g}(\vec{k},\omega_n) = \frac{\tilde{\omega}\underline{\tau}^0 + \tilde{\xi}_k\underline{\tau}^3 + \tilde{\Delta}_k\underline{\tau}^1}{\tilde{\omega}^2 - \tilde{\xi}_k^2 - |\tilde{\Delta}_k|^2} \qquad (1)$$

where the $\underline{\tau}^i$ are the Pauli matrices and $\tilde{\Delta}_k$ is a renormalized unitary order parameter in particle-hole and spin space. The renormalized quantities are given by $\tilde{\omega} = \omega - \Sigma_0(\omega)$, $\tilde{\xi}_k = \xi_k + \Sigma_3(\omega)$, and $\tilde{\Delta}_k = \Delta_k + \Sigma_1(\omega)$, where the self-energy due to s-wave impurity scattering has been expanded $\underline{\Sigma} = \Sigma_i \underline{\tau}^i$. The relevant self-energies are given in a self-consistent $t$-matrix approximation [19,20] by

$$\Sigma_0 = \frac{\Gamma G_0}{c^2 + G_1^2 - G_0^2}; \quad \Sigma_1 = \frac{-\Gamma G_1}{c^2 + G_1^2 - G_0^2}, \qquad (2)$$

where $\Gamma \equiv n_i n/(\pi N_0)$ is a scattering rate depending only on the concentration of defects $n_i$, the electron density $n$, and the density of states at the Fermi level, $N_0$, and we have defined $G_\alpha \equiv (i/2\pi N_0)\Sigma_k Tr[\underline{\tau}^\alpha \underline{g}]$. The strength of an individual scattering event is characterized by the cotangent of the scattering phase shift, $c$. The Born limit corresponds to $c \gg 1$, so that $\Sigma_0 \simeq \pm\Gamma_N G_0$, while the unitarity limit corresponds to $c = 0$. We have defined the normal-state impurity scattering rate as $\Gamma_N \equiv \Gamma/(1+c^2)$; note that in the high-$T_c$ cuprates the total scattering rate at $T_c$ includes inelastic scattering and is expected to be much larger for clean samples.

A crucial feature of the physics of d-wave superconductors is that an infinitesimal concentration of impurities produces a finite density of states $N(0) > 0$ at the Fermi level, [11] leading to temperature dependences characteristic of the normal state in all transport quantities. Solving the self-consistency equations at $\omega = 0$ for the extended-s wave state under consideration leads immediately to the conclusion that such "gapless" behavior is possible only for a range of scattering rates $\Gamma < \Gamma_c$, however. [3] This is illustrated in Figure 3. A low frequency expansion in the gapless regime yields $\Gamma_c/T_{c0} \simeq \eta_0(1+c^2)$.

Can the same experiments which seem to fit the "dirty d-wave" scenario also be explained by extended-s states? The difficulty is how to fix the actual impurity scattering rate, $\Gamma$, given the known concentration. One way is to attribute the additional extrapolated $T \to 0$ resistivity to impurity scattering, such that the elastic and inelastic rates add incoherently. If one attempts such an analysis for Zn-doped YBCO crystals using results from, e.g., Chien et al., [21] one finds that $\Gamma/T_c \simeq 0.3 - 0.5$ per 1% Zn. Since Zn doping studies of YBCO indicate gapless behavior up to several per cent Zn, it would appear that an extended-s picture is plausible for YBCO only if one assumes $\eta_0$ close to the critical value $2/\pi$ for which $\langle \Delta_k \rangle = 0$. [22] We are not aware of similar doping studies in single crystal BSCCO, but assuming for the moment that Zn scatters equally strongly in this material, we see from the dashed line in Figure 3 that a well-developed gap in the excitation spectrum should be induced in BSCCO by a few per cent Zn doping. Note that the results shown in Figure 3 are not sensitive to changes in the scattering phase shift $\delta_0$.

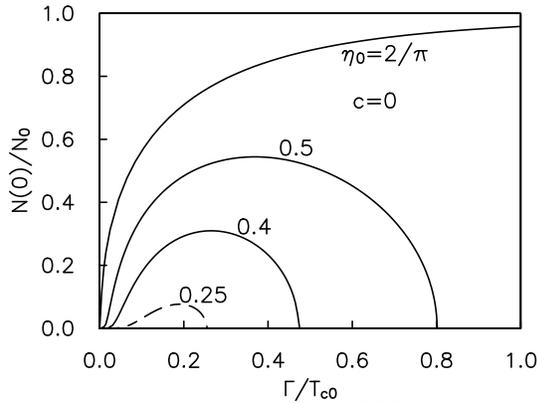

FIG. 3. Residual density of states $N(0)/N_0$ in extended-s state vs. normalized scattering rate $\Gamma/T_{c0}$ for $c = 0$. Dashed line: $\eta_0 = 0.25$ obtained from fit to data of Ding et al.[1]

*Transport properties.* Electrical and thermal conductivities, as well as sound attenuation, will be considerably different in extended s– and d-wave states, as suggested by the following argument. Any DC transport coefficient $L(T)$ in a system characterized by well-defined single-particle excitations will vary with temperature roughly as $L(T) \sim N(\omega \simeq T)\tau(\omega \simeq T)$, where $N(\omega)$ is the density of states and $\tau(\omega)$ the relaxation time. In the clean d-wave case, resonant scattering gives $\tau^{-1}(\omega) = 2\Sigma_0''(\omega) \sim N(\omega)^{-1}$ up to logarithmic corrections, yielding $L(T) \sim T^2$. In the Born limit, $c \gg 1$, similar arguments yield $L \sim const.$ for d-wave transport coefficients.

Impurity-limited transport in the extended s-wave state will be qualitatively similar to the d-wave case if the scattering is weak, $c \gg 1$. In contrast to the d-wave case, however, for $c \to 0$ (and $\eta_0 \ll 1$) resonant scattering does not occur, since the denominator of the t-matrix, $c^2 - G_0^2 + G_1^2 \gtrsim (1-\eta_0)^2$. [4,5] A simple low-$T$ estimate accounting for nodal quasiparticle contributions gives $L \sim L_0(1 - 2\eta_0)$ as $T \to 0$. The exact behavior in this range will be influenced by self-consistency effects and the leading frequency dependence of the t-matrix. The resultant temperature dependence will then be intermediate between the strong and weak scattering limits of d-wave transport coefficients.

*Example: microwave conductivity of extended-s superconductor.* In Ref. 23, expressions for the complex conductivity of an anisotropic s-wave superconductor were derived. We do not reproduce these rather lengthy expressions here, but merely comment that a fully self-consistent numerical evaluation confirms the qualitative picture of low-frequency transport described above. Some typical results are shown in Figure 4 for a model in which inelastic scattering has been neglected entirely. The limiting conductivity as $T \to T_c$ is therefore the impurity Drude result, $\sigma_0 = ne^2/(2m\Gamma_N)$, which for $\Gamma_N \ll T_c$ is much larger than the actual conductivity in the cuprates at the transition, $\sigma(T_c)$, indicated roughly in the figure. At low temperatures $T \ll T_c$, inelastic scattering may be neglected and the results displayed are valid. The most important qualitative feature of the results is that for $\eta_0 < 2/\pi$, the effective limiting value of the conductivity in the extended-s state is nonzero and generically much larger than $\sigma(T_c)$ in clean systems, such that $\bar{\sigma} \equiv \sigma(T \to 0)/\sigma(T_c) \gg 1$. This residual conductivity diminishes as $\eta_0 \to 2/\pi$, when the result should be qualitatively similar to the d-wave conductivity with resonant scattering due to the vanishing of the off-diagonal self-energy. Note, however, that for generic values of $\eta_0$, e.g. that apparently appropriate for BSCCO, *the residual conductivity scales inversely with the impurity scattering rate* $\Gamma$, in contrast to the resonant d-wave case where the residual conductivity is independent of $\Gamma$ to leading order.

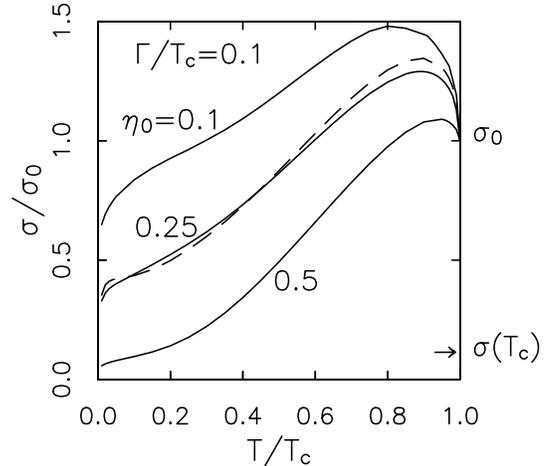

FIG. 4. Conductivity of extended-s state, $\Delta_0/T_c=2.92$, $\Gamma/T_c=0.1$, $c = 0$, $\eta_0=0.1, 0.25$, and 0.5 (solid lines). Also shown, $\Gamma/T_c = 0.01$, $\eta_0 = 0.25$ (dashed line).

As shown in Figure 4, the temperature dependence of the conductivity for most of the low temperature range may mimic a linear behavior. This is intriguing, given the linear-$T$ conductivity observed in microwave experiments on clean YBCO crystals, [6] but the large residual conductivity predicted (unless $\eta_0$ is close to $2/\pi$) would seem to be inconsistent with recent studies indicating that $\sigma(T \to 0)$ is very small in twin-free samples. We are not aware of similar experiments on high-quality BSCCO single crystals. One final point of interest is that the *form* of $\sigma(T)$ is only weakly dependent on disorder, as shown in the figure; of course the overall conductivity scale $\sigma_0$ depends inversely on impurity concentration.

To illustrate the comparison of these results with those expected for a d-wave superconductor, we plot in Figure 5 the conductivity in a $d_{x^2-y^2}$ state in the resonant ($c = 0$) and Born ($c = 10$) limits, together with an example of intermediate strength scattering ($c = 1$) chosen to give the same effective $\sigma(T \to 0)$ as the extended-s conductivity for $\eta_0 = 0.25$. Even in the last case the qualitative differences between the s- and d-wave results are manifest. The dependence of the s-wave result on the scattering phase shift and impurity concentration are found to be quite weak, except in the case $c \gg 1$, for which the result is qualitatively similar to the d-wave Born result, since the denominator of the t-matrix becomes irrelevant.

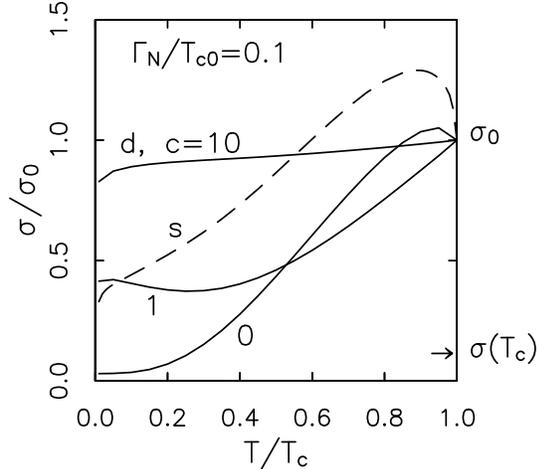

FIG. 5. Comparison of conductivity of clean extended s-wave state (dashed line), $\eta_0 = 0.25, \Delta_0/T_c = 2.92$, $\Gamma/T_c = 0.1, c = 0$ with $d_{x^2-y^2}$ state, $\Delta/T_c = 2.14$, $\Gamma/T_c = 0.1$, $c = 10, 1$, and $0$.

Given the uncertainty regarding the origin of the residual conductivity in the high-$T_c$ materials it is perhaps useful to give estimates for other transport coefficients. For example, since the coherence factors are essentially the same for the electronic thermal conductivity $\kappa_{el}/T$ as for $\sigma(T)$, it is straightforward to see that the large residual conductivity in the extended s-wave state implies a large linear-$T$ term in the thermal conductivity of order $\frac{1}{6}(1 - 2\eta_0)\gamma v_F^2 T/\Gamma_N$, where $v_F$ is the Fermi velocity and $\gamma$ is the normal state linear specific heat coefficient. Such a term has not been observed below $2K$ in single crystal BSCCO [24] but there are data at somewhat higher temperatures which are consistent with $\bar{\sigma}_Q \equiv \kappa_{el}(T \to 0)T_c/\kappa_{el}(T_c)T \gg 1$. [25] The linear term in $\kappa$ at low $T$ in YBCO appears to be quite small. [26]

*Conclusions.* We have shown that, although gapless behavior in thermodynamic quantities qualitatively similar to d-wave states is to be expected for a range of impurity concentrations in extended-s states, transport properties are quite different in the two states. In particular, residual $T \to 0$ conductivities ($\sigma$ and $\kappa/T$) are expected to be large and to scale inversely with impurity concentration, in contrast to the resonant d-wave case. Such experiments can therefore be used in conjunction with Josephson measurements to settle the question of order parameter symmetry. Existing transport data on YBCO single crystals appear to restrict possible s-states to those where the average of the order parameter over the Fermi surface is nearly zero, whereas the BSCCO-2212 material may be consistent with a large average value ($\eta_0 = 0.25$). Further measurements on the latter system, particularly systematic doping studies to search for an impurity-induced gap and to test impurity scaling of residual conductivities, will be necessary before the ARPES identification of an extended-s symmetry order parameter [1] can be proven. A careful determination of the temperature dependence of $\kappa$ over the $1K - 30K$ range in BSCCO would be of great value in this regard.

*Acknowledgement.* Numerical work was perfomed on the Cray YMP at Florida State University.

4